\documentclass{eas}
\usepackage{natbib}
\usepackage{graphicx}
\usepackage{MR_eas}

\begin{document}

\title{MHD simulations of the solar photosphere}
\runningtitle{MHD simulations of the solar photosphere}

\author{M. Rieutord$^1$}
\author{F. Rincon$^1$}
\author{T. Roudier}
\address{Universit\'e de Toulouse; UPS-OMP; IRAP; Toulouse, France
and CNRS; IRAP; 14, avenue Edouard Belin, F-31400 Toulouse, France}

\begin{abstract}
We briefly review the observations of the solar photosphere and pinpoint
some open questions related to the magnetohydrodynamics of this layer of
the Sun. We then discuss the current modelling efforts, addressing among
other problems, that of the origin of supergranulation.
\end{abstract}

\maketitle

\section{Introduction}

The solar photosphere is the only place in the Universe where we have a
detailed view of stellar convection. Thermal convection is a well-known
phenomenon that has been studied for more than a century, but in the case
of stars it still owns many dark sides that prevent us from a full
understanding. Unfortunately, convective regions of stars like the Sun are
the seat of the magnetic activity, whose explanation requires strong
investigations of flows where buoyancy, radiation and magnetic fields
couple together.

Flows are thus complex, but their darkest side is their turbulent
nature which implies the interaction between many scales either in the
velocity field or in the magnetic field or between both.  Handling such
a multiscale phenomenon has become possible when computers have reached
enough computing power so that numerical simulations be realistic on
some side(s). The first work of \cite{nordlund85} perfectly illustrates
the emergence of interesting simulations coupling fluid dynamics
and radiative transfer, and which could be compared to observations
(line profiles). With the steady increase of computational power, such
simulations have become the preferred tool of astrophysicists involved
in fluid dynamical problems. More and more sophisticated simulations
addressing the solar photosphere magnetohydrodynamics have emerged
\cite[][]{SGSNB09,ustyugov09}.

Thus, in this short review, we first set up the stage provided by
observations of the Sun and pinpoint some open questions. Then, we
briefly describe the current modelling efforts, ending this work with some
perspectives.

\section{Observations of the quiet Sun dynamics}
\subsection{Multiscale convection}

It is well-known that solar convection is a multiscale phenomenon: first
of all because it is a turbulent flow that naturally contains a
continuum of scales and second because of some specifities of thermal
convection in stars that single out some particular scales.

\begin{figure}
  \centering
  \resizebox{0.98\hsize}{!}
   { \includegraphics{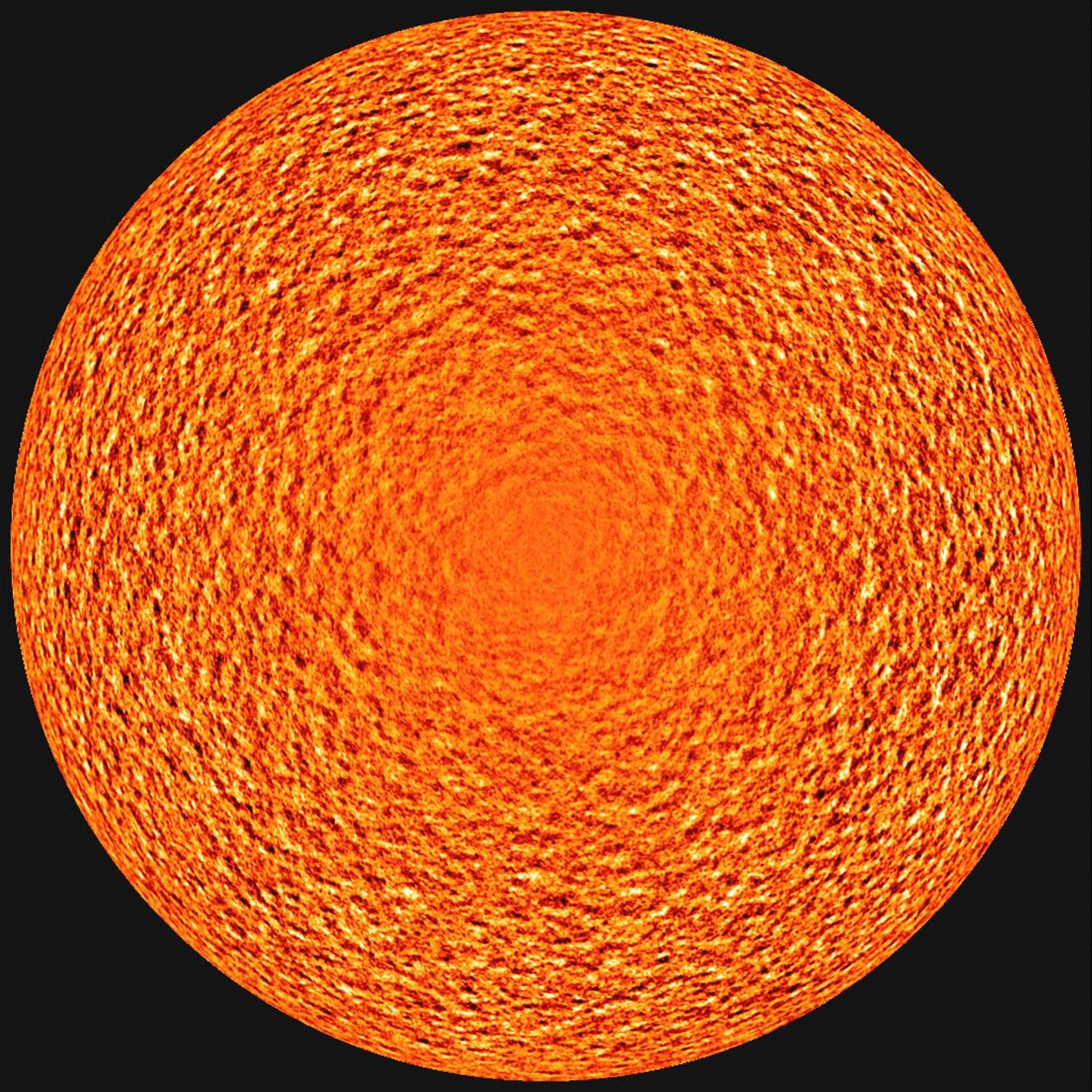} }
  \caption[]{\label{dopplergram}The solar supergranulation as seen through a
dopplergram (credit SOHO/MDI/ESA).}
\end{figure}

Among the specific scales, the most well-known is the granulation,
discovered long ago by Nasmyth in 1860 \cite[][]{bartholomew76} and which
shows cells typically 1~Mm wide, lasting 500-1000s with velocities
of order 1-2~km/s. As detailed below the physical origin of granulation
is understood. This is not quite the case of the supergranulation which
is the other specific scale arising in the convective flow visible at
the Sun's surface. This scale has also been known for quite some time,
exactly since the work of \cite{hart54}. Unlike granulation, it is only
visible via its horizontal velocity signature. Thus, dopplergrams like
that of Fig.~\ref{dopplergram} readily show this feature. The typical
length scale is around 30~Mm, with a time scale typically 1.8 days,
meaning velocities about 400~m/s (we refer the reader to \citealt{RR10} for
a detailed review of solar supergranulation). For historical reasons, we only
briefly mention mesogranulation, which has been searched for a long
time as the intermediate scale between granulation and
supergranulation. Results have been much controversial and we think
that mesogranulation is most likely a ghost pattern emerging from data
processing of delicate flow measurements \cite[see][]{RRMR00,RRRMMBZ10}.

\subsection{Magnetic fields}

\begin{figure}
  \centering
  \resizebox{\hsize}{!}
   { \raisebox{15mm}{\includegraphics{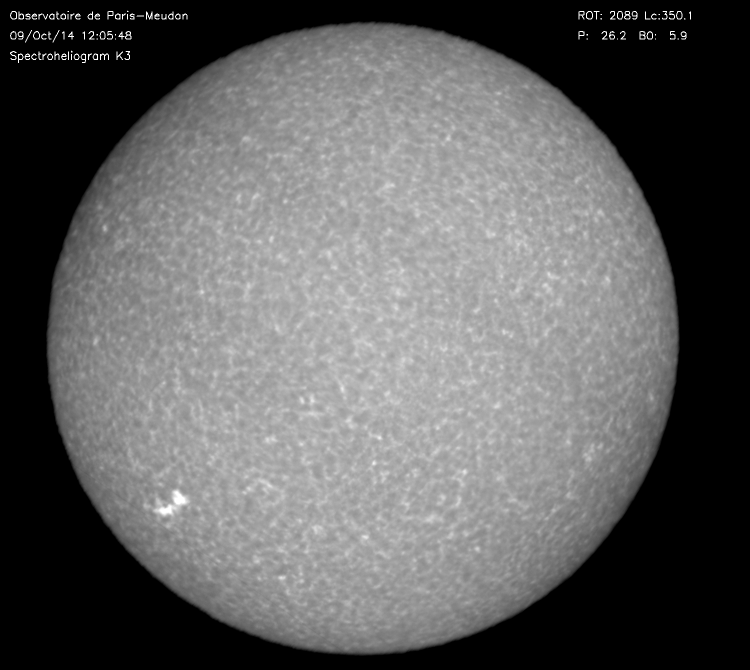} }
    \hfill\includegraphics{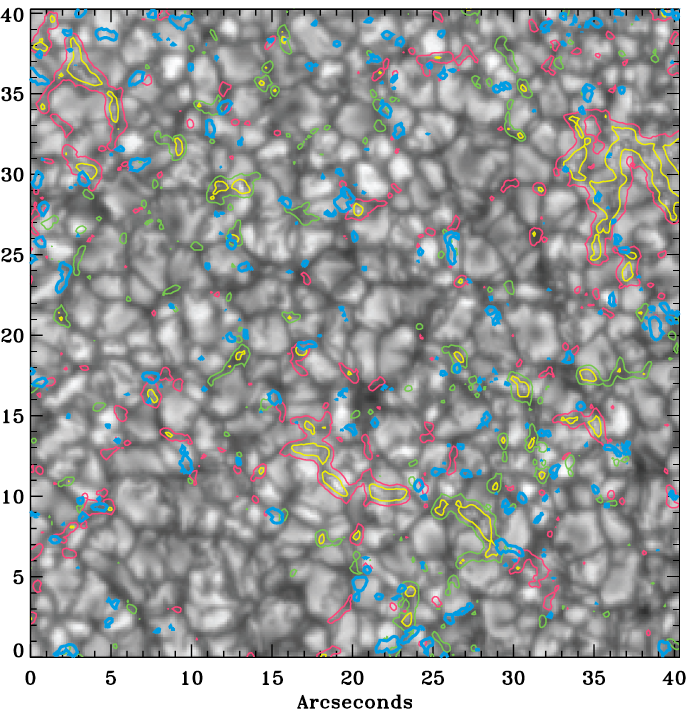} }
  \caption[]{\label{inb_nb}Left: the network field as seen through the
chromospheric line Ca+ K3 line (credits SOHO/MDI/ESA). Right: the intra
network fields seen with Hinode \cite[from][]{Lites_etal08}.}
\end{figure}

In the quiet photosphere magnetic fields are also present especially
through the so-called ``network" which is clearly seen with the
chromospheric Ca$^+$ K3 line. The network shows a cellular pattern covering
the Sun (see Fig.~\ref{inb_nb}), and coincident with the supergranulation
cell boundaries, but with slightly smaller cells 
\cite[e.g.][]{HST97}. It has thus often been used to monitor the size
variations of supergranulation.  The origin of this field, which consist
of flux tubes with a magnetic field up to 1kG, is not fully understood,
especially its interaction with supergranulation. The situation is quite
similar for the intra-network field, which is very disordered and with
an amplitude in the range 50-200~G \cite[][]{dominguez03}.

Recently, \cite{roudier_etal09} studied the evolution of ``Trees of
Fragmenting Granules" (TFG) and found that they were basically advected by
the supergranular flow. Besides, using floating corks, they demonstrated
that vertical magnetic fields always evolve to a patchy distribution
along the supergranule boundaries. This enforces the relation between
magnetic fields and supergranulation, suggesting that this feature  is
an emergent length scale building up when small magnetic elements are
displaced by TFGs flow, occasionally colliding and aggregating to form
larger magnetic clusters, which in combination with granulation can
trigger the supergranular downflow structure.

\section{Open questions}

\begin{figure}
  \centering
  \resizebox{\hsize}{!}
   { \includegraphics{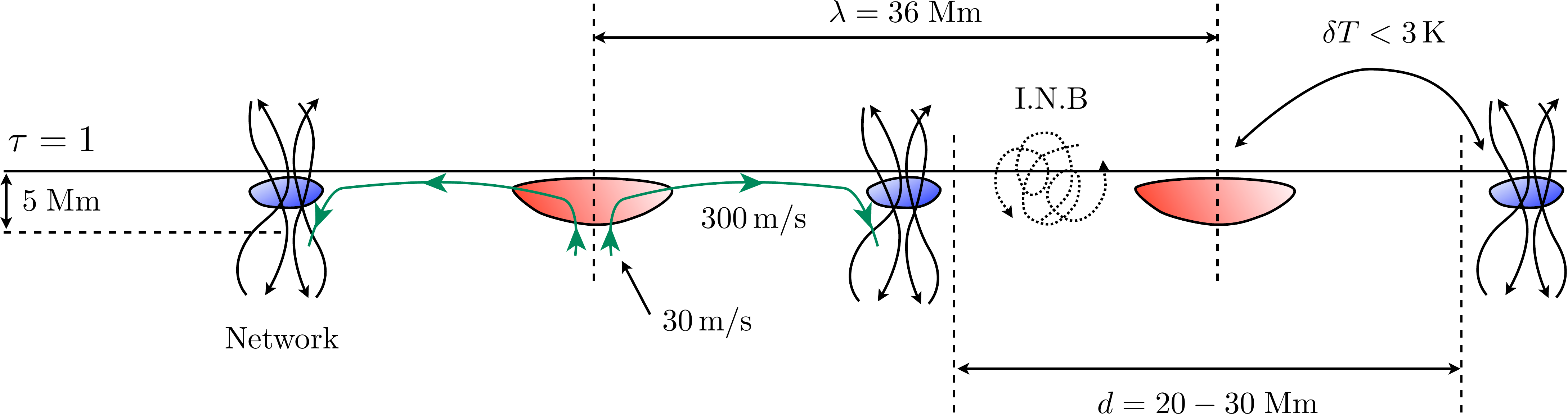} }
  \caption[]{\label{big_pict} A summary of the large-scale flows and magnetic
fields in a quiet region of the Sun \cite[from][]{RR10}.}
\end{figure}

\begin{figure}
  \centering
  \resizebox{\hsize}{!}
    {\includegraphics{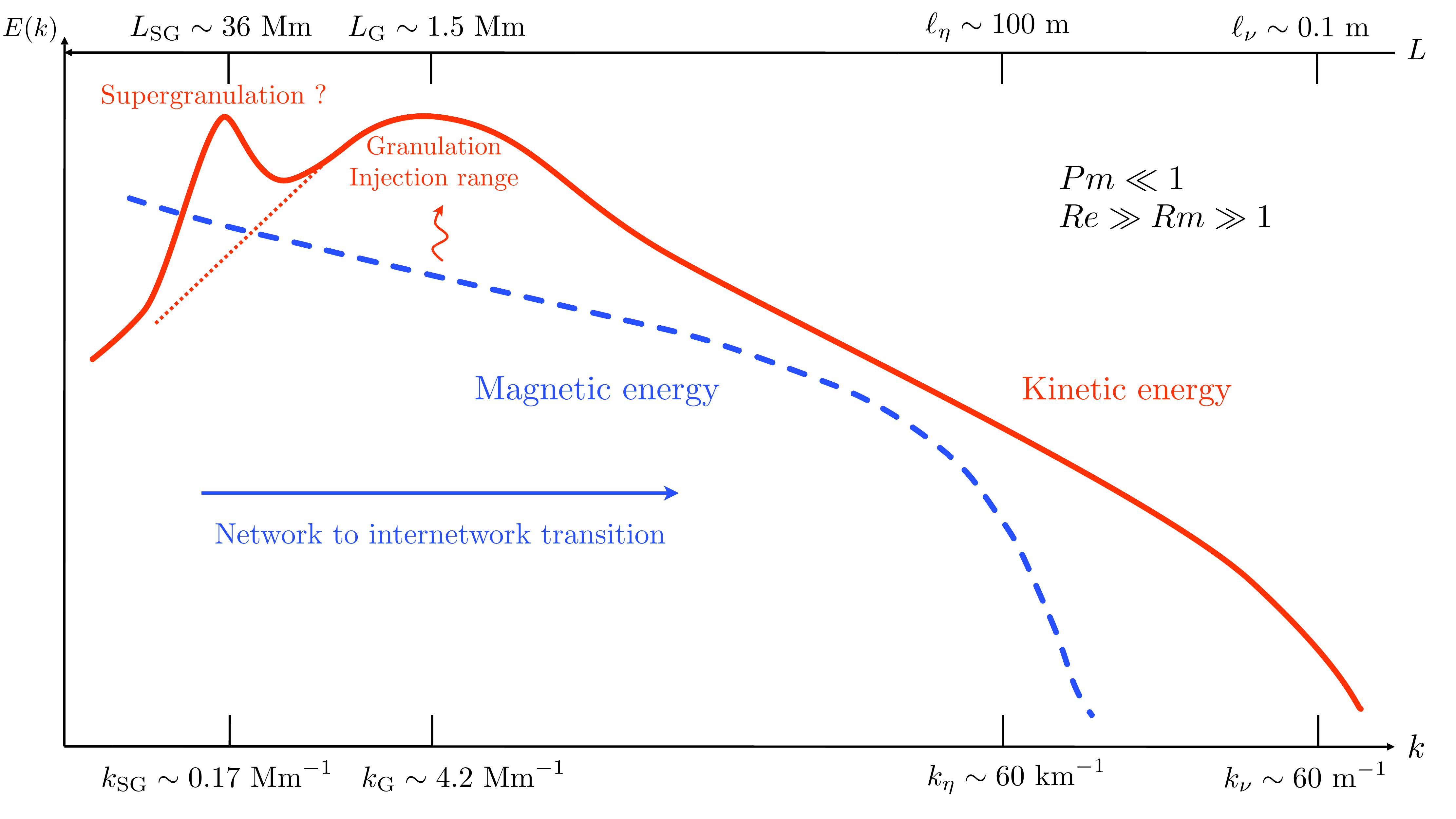} }
\caption[]{\label{big_spec} A schematic view of the kinetic and magnetic
spectra that should be assessed by observations \cite[from][]{RR10}.}
\end{figure}

The observations, very briefly summarized above, raise many unsolved problems.
The main one, but not the only one, is that of the origin of the
supergranulation. The observational facts may be summarized by the
schematic view displayed in Fig.~\ref{big_pict}. Many scenarii have been
proposed to explain the existence of supergranulation \cite[][]{RR10}, but
none of them is confirmed. At the moment, we (the authors) favour the idea
that this feature of solar surface flows results from a large-scale
instability of surface convection, likely influenced by the magnetic
fields, as suggested by Fig.~\ref{big_spec} \cite[see also][]{RR03}.

However, this question is most probably related to the origin of the
intra-network fields: are these fields generated locally by a small-scale
dynamo ? or reprocessed from active regions or just emerging from beneath ?
How do they relate to the cycle ? Can we distinguish these various possible
origins?

%\subsection{Production of magnetic fields in quiet regions}
%\subsection{Large-scale MHD, network and supergranulation}

\section{Current modelling efforts}
\subsection{Solar convection: a very turbulent flow}

To really appreciate the difficulty of the modelling of flows in the solar
photosphere, one should first realize how extreme are the numbers which
control solar convection.

The Rayleigh number is the ratio between buoyancy, which drives the flows,
and dissipative process (viscosity and heat diffusion). In laboratory
experiment, one hardly reaches 10$^{11}$, while the Sun reaches 10$^{22}$.
Reynolds numbers in solar flows are typically around 10$^{12}$, while
laboratory experiment remain below 10$^{9}$. This number is particularly
interesting since it controls the ratio between the smallest (dissipative)
scales and the most energetic ones (the one we see). Indeed this ratio
scales like Re$^{3/4}$. This means that, for instance, granular flows for
which Re$\sim10^{12}$, own structures (vortices) whose size range from
1000~km down to 1~mm!

The extreme values shown by Rayleigh and Reynolds
numbers are essentially due to the large size of the body. However, the
fluid itself has intrinsic properties which are uncommon to Earth
standards. The kinematic viscosity $\nu$ of the fluid is around
$10^{-3}$~m$^2$/s \cite[e.g.][]{R08}, which is a thousand times larger
than that of water. However, the heat diffusivity $\kappa$ is 10$^5$
times higher because heat is carried by photons essentially and radiative
processes are increasingly efficient when temperature raises. The point
here is that the Prandtl number, ${\cal P} =\nu/\kappa \sim 10^{-5}$, which is
very small compared to that of any terrestrial fluid (the lowest value
known on Earth is that of mercury which is 0.025). The same occurs for the
magnetic field diffusivity, controlled by the electric conductivity of the
plasma. The magnetic Prandtl number is also very small, in the range
10$^{-5}$--10$^{-2}$.

The values of all these numbers make a direct numerical simulation of solar
convection strictly impossible with nowadays computers. Likely, this will
remain the case for many many years, and actually such a DNS would not be
so interesting except of being an experiment where one can play with huge
statistics on all sorts of quantities!

\subsection{Current simulations}

\subsubsection{Granulation}

Nowadays simulations of the solar photosphere
\cite[e.g.][]{SN98,RLRNS02,SGSNB09} do not take into account the
very high values of the Rayleigh and Reynolds numbers. Their viscosity are
actually of numerical origin and related to the mesh size. They thus belong
to the category of Large Eddy Simulations (LES), which use artificial
truncation of the turbulent spectrum.

These LES have however been able to reproduce the thermal structure of the
surface layers of the Sun and especially granulation \cite[we refer the
reader to][for a very recent comparison of the codes]{beeck_etal12}. The
very reason for this success is that the scale of granulation is such that the
P\'eclet number is of order unity. It means that heat diffusion and heat
advection are of the same order of magnitude, and therefore heat transport
is correctly taken into account. Of course the Reynolds number is
unrealistic: actually, a visual comparison between the simulated and
observed granules shows that the latter ones are much more turbulent and
slightly (by 10-15\%) smaller (Roudier, private communication).

\subsubsection{Supergranulation}

The success obtained in modelling granules has triggered attempts to
simulate supergranulation \cite[][]{RLRNS02,SGSNB09}. Unfortunately, these
attempts failed at exhibiting this feature of the solar convection. There
are likely several reasons for that failure. The first one is the
increased (over granulation) difficulty set up by the necessary size
of the computing domain. The aspect ratio (width over depth) has to be
increased sufficiently to give room to this structure. If we assume,
which is not proved, that supergranulation is a surface phenomenon (like
granulation), then computing boxes just need to be extended horizontally,
ideally by a factor of order 30 (the ratio between supergranular and
granular scale), meaning a factor 900 on the grid points. None of the
simulations have made such a jump, therefore the granules computed
in these supergranulation-intended-simulations face an even greater
numerical viscosity. Thus, if supergranulation comes form a large-scale
instability of sub-structures, the reduced Reynolds number of the
sub-structures (like granule), is likely to impede the development
of such an instability. In addition, such big boxes require much more
time to relax. As a consequence, if the simulations is on the verge of
displaying the instability, the supergranulation pattern might not be
seen just because the computation is not long enough!

\subsubsection{Magnetic fields}

\begin{figure}
  \centering
  \resizebox{\hsize}{!}
    {\includegraphics{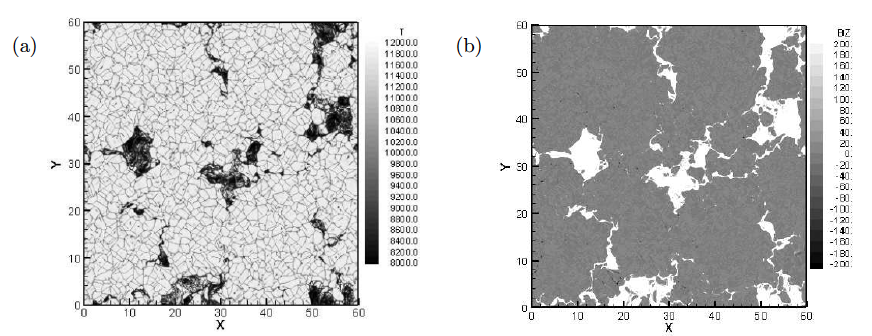} }
\caption[]{\label{network} View of the temperature field (left) and
vertical magnetic (right) as obtained by \cite{ustyugov09}; the X and Y
scales are in Mm.}
\end{figure}

In the previous simulations no magnetic field is included, however it might
be crucial for the existence of supergranulation. Including magnetic fields
is costly and therefore attempts in this direction have relaxed other
constraints, essentially the size of the domain. For instance,
\cite{ustyugov09} includes magnetic fields but uses a resolution almost
twice smaller than \cite{SGSNB09}. In addition, the run lasts a shorter
time (48 solar hours instead of 64 s.h.). Finally, in this simulation,
the magnetic field was not generated by a local dynamo, but given with
initial conditions (an initial uniform vertical field). The results
are nevertheless interesting as a structure like the network appears (see
figure~\ref{network},
however the simulation is not long enough to tell whether this structure
is statistically steady. Obviously, more efforts are needed in this
direction.

 Beside the large values of the kinematic and magnetic Reynolds numbers,
another difference between simulations and actual stellar situations is the
ratio between these two numbers, namely the magnetic Prandtl number $\PRM$.
Stellar values are small compared to unity, while simulations have
displayed a dynamo when using either order unity values for $\PRM$
\cite[][]{BJNRST96} or values larger than unity
\cite[][]{NBJRRST91,cattaneo99}. At Reynolds numbers reachable by
simulations, the dynamo disappears when the magnetic Prandtl number is
realistic. 

\begin{figure}
  \centering
  \resizebox{0.8\hsize}{!}
    {\includegraphics{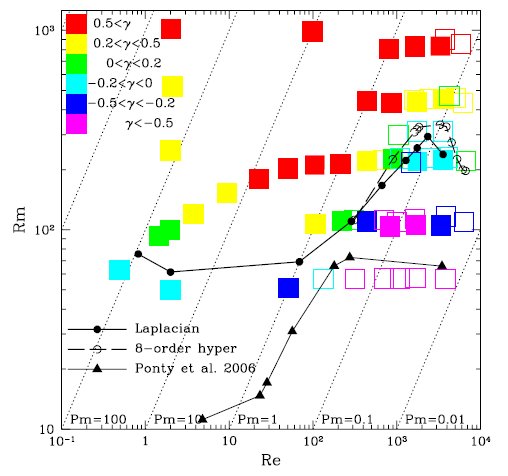} }
\caption[]{\label{schek} Positions of kinematic dynamo simulations
by \cite{schekochihin_etal07} using viscosity or hyperviscosity as a
representation of the neglected small scales. The solid lines give the
critical magnetic Reynolds numbers according to this prescription or from
the work of \cite{ponty_etal07}.}
\end{figure}

Actually, it seems that the critical value of the magnetic Reynolds
number increases when the kinetic Reynolds number increases. This
effect is expected since a higher kinetic Reynolds numbers means
a stronger turbulence and thus a more effective dissipation. This
effect does not facilitate numerical investigations. Recent work
by \cite{schekochihin_etal07} investigated a kinematic dynamo at low
magnetic Prandtl number (down to $\PRM=0.07$), showing that velocity
field and the magnetic have very different statistical properties.
Figure~\ref{schek} from \cite{schekochihin_etal07} shows the positions
of several dynamo simulations in a diagramme (Re, Re$_m$), delineating
the critical line below which the dynamo instability disappears. Note
that for this kind of flows, this critical curve Re$^{\rm crit}_m$(Re)
is not monotonous.
Another recent work by \cite{buchlin11}, used the simplified shell model
and argued that the critical Magnetic Reynolds number remains finite
as the kinetic Reynolds number tends to infinity. However both of the
foregoing works are investigating kinematic dynamos and presently,
nobody knows how such low-$\PRM$-dynamos saturate.

\section{Perspectives}

To conclude this short review, we would like to insist on the problems
faced by Large-Eddy Simulations (LES) at simulating the solar photosphere.

Obviously, LES are still lacking of a good subgrid scale model to represent
more faithfully the effects of small-scales that are not resolved by the
simulations. Actually, this is a case of fundamental research in
turbulence, which is much investigated in applied fluid mechanics, but not
so much in astrophysics. The astrophysical situations are of course not
amenable to laboratory experiments, but the numerous detailed observations
that can be gathered from the Sun are very useful to give hints to models.
For instance a determination of the magnetic energy spectrum completing the
work of \cite{abramenko01}, may be very helpful to find the origin of
supergranulation.

Back to simulations of the photosphere, more detailed investigations are
needed to understand the interplay of a small-scale dynamo, its saturation
at low magnetic Prandtl number, and the basic advection of a background
field. These are likely questions for the next decades...

%\bibliographystyle{aa}
%\bibliography{../../biblio/bibnew}

\end{document}